\documentclass{iau} 
\usepackage{graphicx}
\usepackage{subfigure}

\title[Energy distribution of solar flare events] 
{Energy distribution of solar flare events}

\author[S. Sen; A. Mangalam \& R. Ramesh]   
{S. Sen$^1$, A. Mangalam$^2$ \and R. Ramesh$^3$}

\affiliation{Indian Institute of Astrophysics\\ Sarjapur Road, Koramangala, Bangalore, 560034, India, \\ email: {\tt $^1$samrat@iiap.res.in, $^2$mangalam@iiap.res.in, $^3$ramesh@iiap.res.in}} 

\pubyear{2018}
\volume{340}  
\jname{Long-term datasets for the understanding of solar and stellar magnetic cycles}
\editors{A.C. Editor, B.D. Editor \& C.E. Editor, eds.}
\begin{document}

\maketitle

\begin{abstract}
Observational evidence of the braiding of magnetic field lines has been reported. The magnetic reconnection within the loop (nanoflares) and with other loops (microflares) disentangle the field. The coronal field then reorganizes itself to attain a force-free field configuration. We have evaluated the power law index of the energy distribution $f(E)=f_0 E^{-\alpha}$ by using a model of relaxation incorporating different profile functions of winding number distribution $f(w)$ based on braided topologies. We study the radio signatures that occur in the solar corona using the radio data obtained from the Gauribidanur Radio Observatory (IIA) and extract the power law index by using the Statistic-sensitive nonlinear iterative peak clipping (SNIP) algorithm. We see that the power law index obtained from the model is in good agreement with the calculated value from the radio data observation.

\end{abstract}

\firstsection 
\section{Introduction}
\begin{figure}
\begin{center}
\includegraphics[scale=0.3]{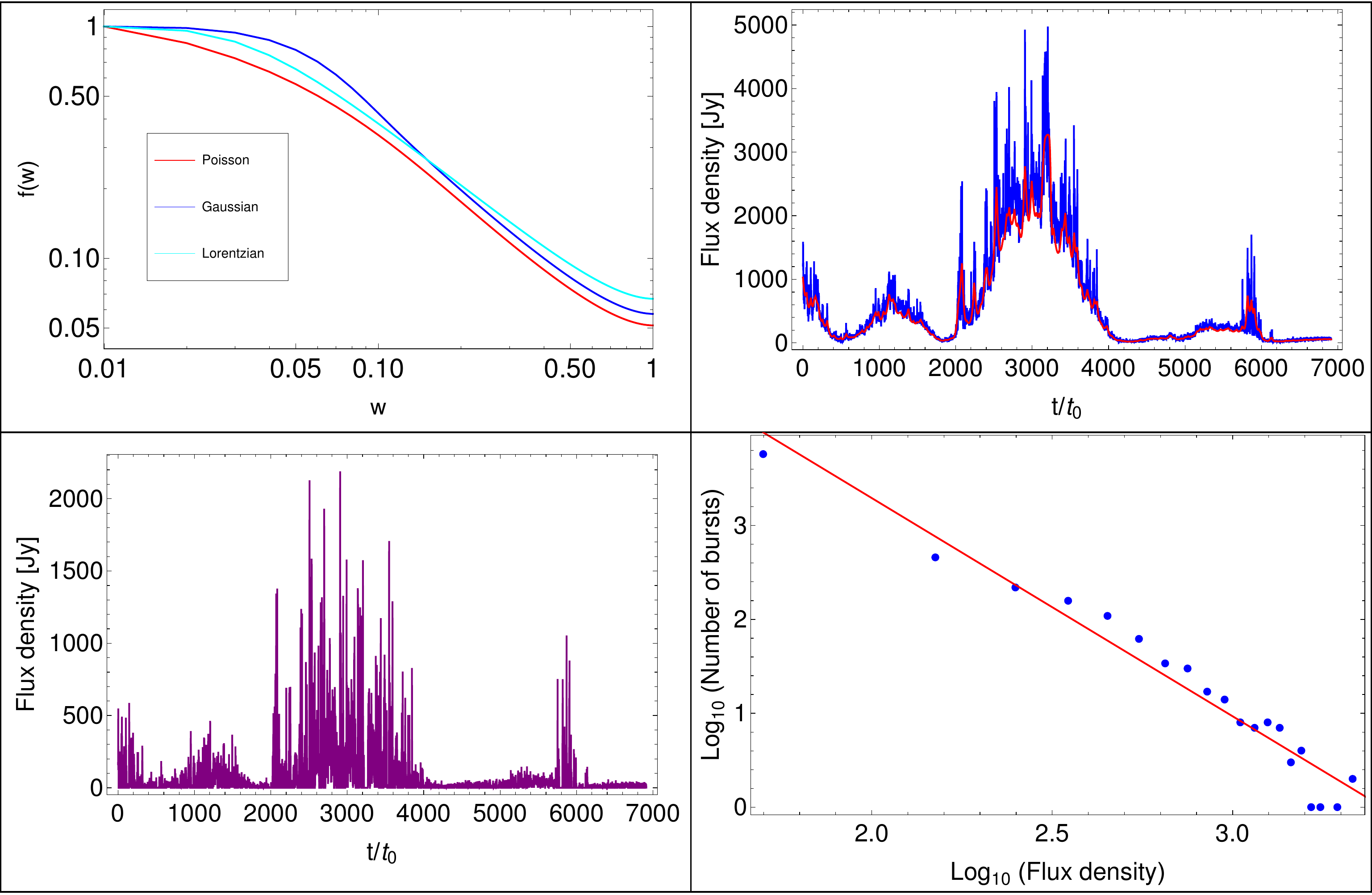}\label{plots}
\caption{(Upper left) The distribution of the winding number, $w$ for Poisson, Gaussian and Lorentzian profiles. (Upper right) The radio signature of the solar flare. The time axes in the figures on the  upper right and lower left are scaled with $t_0=500$ ms. The  figure in the lower right is the logarithm fitting of the flux density vs burst frequency plot. The value of the slope is $-2.51$.}  
\end{center}
\end{figure}

An analytic model of braided magnetic fields which gives a power law distribution of energy releases is presented in \cite[Berger \& Asgari-Targhi (2009)]{Berger_Targhi2009}. \cite[Mangalam \& Prasad (2018)]{Mangalam_Prasad2018} have incorporated nonlinear force free equation (\cite[Prasad et al. 2014]{Prasad_etal2014}; \cite[Prasad \& Mangalam 2013]{Prasad_Mangalam2013}) to calculate the winding numbers, linkages to derive an analytical bound of the free energy and relative helicity. The braiding of the different field lines to each other is called the coherent sequence, and the swapping of the different field lines is called the interchange. In order to model the self-organized criticality (SOC), \cite[Berger \& Asgari-Targhi (2009)]{Berger_Targhi2009} have considered a braiding system having $m$ number of sequences and $m-1$ number of interchanges. At each evaluation step, one new coherence sequence is added with one interchange, and simultaneously a reconnection event eliminates one of the existing interchanges. If $n(w)$ is the sequence having crossing number $w$ and $f(w)$ be the probability distribution function of sequence length $w$, then at each time step the  change in $\delta w$ corresponds to the change in the $n(w)$ is $\delta n(w)$. Taking the Fourier transform of the relaxed state i.e. $\delta n(w)=0$, we obtain the relation, $\tilde{f}(k)=1-\sqrt{1-\tilde{p}(k)}$, where $\tilde{f}(k)$ and $\tilde{p}(k)$ are the Fourier transforms of $f(w)$ and $p(w)$ respectively. \cite[Berger \& Asgari-Targhi (2009)]{Berger_Targhi2009} have estimated the $f(w)$ distribution by taking $p(w)$ to be the Poisson distribution, ${\displaystyle p_P(w)=\frac{\lambda}{2} \exp(-\lambda | w |)}$. We have extended the model by incorporating two different profile functions:  Gaussian, ${\displaystyle p_G(w)=\frac{1}{\lambda\sqrt{2\pi}}\exp\bigg(\frac{-w^2}{2\lambda^2}\bigg)}$ and Lorentzian, ${\displaystyle p_L(w)=\frac{\lambda}{\pi(\lambda^2+w^2)}}$ into the model. We evaluate the distribution of $f(w)$ for all the profiles, shown in the upper left panel of Figure \ref{plots}. The reconnection between the field lines occur when the crossing number reaches to a certain critical limit. Assuming the coherence length follows the power law $f(w)=b |w|^\beta$, the energy distribution follows the relation, $F(E) \propto E^{-\alpha}$, where $\alpha=2\beta-1$. The value of the $\beta$ is obtained by taking the average value of $\displaystyle{\frac{{\rm d}\ln(f)}{{\rm d}\ln(w)}}$ over $w$, and hence the value of $\alpha$ is estimated to be $2.95$, $2.5$ and $0.94$ for the Poisson, Gaussian and Lorentzian profiles respectively. 
We use the radio data obtained from Gauribidanur Radio Telescope (\cite[Ramesh 2011]{Ramesh2011}; \cite[Ramesh et al. 2013]{Ramesh_etal2013}) at $80$ MHz frequency which was taken on 12 March, 2011. The data corresponds to the temporal distribution of the Sine and Cosine visibilities. We take the square root of the sum of the squares of the visibilities which gives the antenna response. Now, we eliminate the background noise by the Statistic-sensitive nonlinear iterative peak clipping (SNIP) algorithm (\cite[Tomoyori et al. 2015]{Tomoyori_etal2015}), shown in the upper right panel of Figure \ref{plots}, where the blue curve is the antenna response and the red curve is the background noise fitting. The temporal distribution of the flux response without noise is shown on the lower left panel of Figure \ref{plots}. The figure in the lower right panel of Figure \ref{plots} shows the logarithm distribution of the frequency of the radio bursts vs flux density. We fit a linear curve with the logarithmic distribution of the flux density vs number of radio bursts, and find the slope to be $-2.51$. 

\vspace{-0.2 in}
\section{Conclusions}
The value of the slope depicts that the power-law index, $\alpha$, of the energy distribution, $f(E)=f_0 E^{-\alpha}$ is $2.51$. This value is in good agreement with our theoretical predictions for Poisson and Gaussian profiles which are $2.95$ and $2.5$ respectively, but not for the Lorentzian case where the value is $0.94$.

\end{document}